\documentstyle[12pt,aps]{revtex}
\begin{document}

\begin{titlepage}
\thispagestyle{empty}
\begin{center}
\vspace*{1cm}
{\bf \Large Exclusive Many-Particle Diffusion in Disordered Media
and Correlation Functions for Random Vertex Models}\\[23mm]

{\Large {\sc
Gunter Sch\mbox{\"u}tz\footnotemark[1] \vspace*{2mm} \\
and \vspace*{4mm}\\ Sven Sandow}\footnotemark[2]
} \\[5mm]

\begin{minipage}[t]{13cm}
\begin{center}
{\small\sl
  \footnotemark[1]
  Department of Physics,
                Weizmann Institute, \\
  Rehovot 76100,
  Israel
  \\[2mm]
  \footnotemark[2]
  Department of Electronics,
                Weizmann Institute, \\
  Rehovot 76100,
  Israel}
\end{center}
\end{minipage}
\vspace{20mm}
\end{center}
{\small
We consider systems of particles hopping stochastically
on $d$-dimensional lattices with space-dependent probabilities.
We map the master equation
onto an evolution equation in a Fock space where the dynamics
are given by a quantum Hamiltonian (continuous time) or a transfer
matrix resp. (discrete time). We show that under certain conditions
the time-dependent two-point density
correlation function in the $N$-particle steady state
can be computed from the probability distribution of a single
particle moving in the same environment.
Focussing on exclusion models where each lattice site
can be occupied by at most one particle we discuss
as an example for such a stochastic process
a generalized Heisenberg antiferromagnet where the strength
of the spin-spin coupling is space-dependent. In discrete time one
obtains for one-dimensional systems the diagonal-to-diagonal
transfer matrix of the two-dimensional
six-vertex model with space-dependent vertex weights.
For a random distribution of the vertex weights one obtains a version
of the random barrier model describing diffusion of particles in
disordered media. We derive exact expressions for the averaged two-point
density correlation functions in the presence of weak, correlated
disorder.
}
\\
\vspace{5mm}
\end{titlepage}

\newpage

\section{Introduction}
\renewcommand{\theequation}{\arabic{section}.\arabic{equation}}
\setcounter{equation}{0}

Stochastically hopping particles on a lattice represent a simple model
for diffusion in various media. Of particular interest in recent years
has been the influence of static disorder on the dynamics.
The diffusion of a single
particle on a lattice with random hopping rates is well understood
\cite{bo} - \cite{hav}. Clearly, a system of interacting
particles
does not behave as a mere superposition of single particles. The
effect of a hard core interaction on diffusion in ordered system has been
intensively studied recently \cite{KDN} - \cite{es}. However
the dynamics of exclusive particles in a random system poses a much more
difficult problem. Some examples have been analyzed numerically \cite{ke}
or analytically in a mean field approximation \cite{sa3}. However,
very few exact results are known.

The main aim of this paper is to derive properties of correlation
functions for certain diffusive systems of exclusive particles
(particles with a hard-core repulsion)
in a disordered environment. We consider $d$-dimensional
models where the
diffusion of particles is determined by the presence of barriers
of different strength. For these models
the problem of calculating time-dependent
(sometimes called ``time-delayed'') correlation functions
in the steady state of the {\em many-particle} system will
be shown to be reducible to a {\em one-particle} problem.

We will use a description by means of stochastic occupation numbers on a
lattice, the dynamics of which are given by a master equation in either
discrete or continuous time. A very effective way of
treating master equations is given by a Fock space method introduced
by Doi \cite{doi}, further extended and applied in
\cite{pa} - \cite{jens}.
In this formalism noninteracting particles are represented as
bosons while the hard core interaction preventing the occupation
of a lattice site by more than one particle
gives rise to a fermionic character of the
particles.

Besides its usefulness for solving various problems, the mapping of the
classical system onto a Fock space opens deeper insight into
the relations to
certain quantum systems. So we find the diffusion of exclusive particles
equivalent to a Heisenberg antiferromagnet.
The correlations of occupation
numbers correspond to spin-spin correlations in the antiferromagnet.
Such a mapping is particularly interesting for one-dimensional models
where the Hamiltonian obtained
for many equilibrium and non-equilibrium
processes such as diffusion-limited chemical reactions or driven
diffusive systems \cite{GS,adhr} turns out to be related to
quantum Hamiltonians of integrable models. Thus part of
the vast amount of knowledge that has accumulated for these models
over the past decade and some of the methods applied for solving them
such as the Bethe ansatz \cite{sch1,GS} are useful
also for the classical systems from which the Hamiltonian was obtained
and leads the way to new predictions.

One obtains quantum {\em Hamiltonians} for processes in continuous
time, but one may also study stochastic processes in discrete
time. In this case the mapping to a Fock space gives rise to a
{\em transfer matrix} which again, for many interesting problems
in one dimension, turns out to be related to the transfer matrix
of integrable two-dimensional systems such as vertex models
\cite{KDN,sch1,sch2,bax}. Here we will utilize this mapping
for a study of correlated disorder in a one-dimensional many-particle
random barrier model with exclusion. In this model the probability $p_x$
that a particle moves from a lattice site $x$ to its right
neighbouring lattice site $x+1$
is the same as for a jump from $x+1$ to site $x$. However $p_x$ is
space-dependent, i.e., different for each pair of sites $(x,x+1)$.
We define a parallel updating mechanism
where these hopping rules are
applied first to all pairs of sites $(2x-1,2x)$ and then in a
next step to all shifted pairs $(2x,2x+1)$ as in Ref. \cite{KDN}.
The discrete time dynamics of this model
are thus encoded in the
diagonal-to-diagonal transfer matrix of a six vertex model
with certain randomly chosen vertex weights.

Much is known about the (continuous-time) single-particle
random barrier model with uncorrelated
disorder, where all hopping rates $p_x$ are independently chosen
random numbers with the same distribution $g(p)$.
The averaged probability distribution
of finding the particle has been studied in detail by many authors
\cite{bo} - \cite{zw}.
For distribution functions where the moments of the inverse
hopping rates do not exist, a phase transition has been
shown to take place \cite{bo,ha,te}. Only a few results are known
about systems with spatially correlated disorder \cite{hav,val}.

We will study the behaviour of the averaged density two-point
correlation function in the steady state of the many particle
system in an environment with correlated disorder. In Section 2
we show under which conditions this correlation function can
be expressed in terms of the one-particle density distribution.
This discussion is quite general and applies to ordered and
disordered models in any dimension.
In Section 3 we apply our result to systems where the corresponding
quantum Hamiltonian turns out to become a generalized Heisenberg
antiferromagnet with spatially dependent strength of the
nearest-neighbour spin-spin coupling. We also briefly review
some results on the one-dimensional case. In section 4 we
present our results on the discrete-time formulation of the
process, i.e., the 6-vertex model. We briefly indicate the
existence of a new type of phase transition in the homogeneous
model with equal and fixed hopping rates $p$ as they approach
the value 1. Furthermore we derive exact expressions for the
averaged correlation function for weak, correlated disorder
and discuss the impact of the correlations on its behaviour as
compared to the uncorrelated case.
In Section 5 we summarize our results and in the
Appendices we present some details of our calculations to
Section 2 (in Appendix A) and we repeat and generalize the
mapping of Ref. \cite{KDN} of the diffusion problem to the
6-vertex model (in Appendix B).

\section{The two-point correlation function for a
many-particle system}
\setcounter{equation}{0}
\newcommand{\BIN}[2]{\renewcommand{\arraystretch}{0.8}
\mbox{$\left(\begin{array}{@{}c@{}}{\scriptstyle #1}\\
{\scriptstyle #2}\end{array}\right)$}\renewcommand{\arraystretch}{1}}

In this section we will show how the two-point correlation function of a
many particle system is related to
the probability distribution of a single particle moving in the same
environment.

The many particle system is assumed to be described by a set of
occupation numbers on a d-dimensional lattice $\underline {n}= \{n_j\}$ .
Its dynamics are given by a master equation of the following form:
\begin{equation}\label{2A}
\partial_t F(\underline{n},t)=H'F(\underline{n},t)
\end{equation}
where $H'$ is some linear operator acting on the $n_{j}$. We may also
consider a discrete time dynamics where the
time derivative in Eq. (\ref{2A}) is replaced by a discrete difference.
Instead of $H'$ a transfer matrix $T'$ is then used (see
Section 4).

According to Doi's formalism the master equation is mapped onto an
evolution equation in a Fock space \cite{doi} - \cite{sa1},
\cite{ru}:
\begin{equation}\label{2B}
\partial_t
|\; F(t) \; \rangle=H|\; F(t) \; \rangle
\end{equation}
or a similar equation in case of a the discrete time. The solution of
Eq. (\ref{2B}) can be written as:
\begin{equation}\label{2C}
|\; F(t) \; \rangle=U_t |\; F(0) \;
\rangle \hspace*{4mm} \mbox{with} \hspace*{4mm}
U_t=\mbox{e}^{Ht}
\hspace*{2mm} .
\end{equation}
(For a discrete time dynamics $U_t$ after $t$ time steps is given by the
$t$-th power $T^t$ of the transfer matrix $T$.)

A particular configuration $\underline{n}$ corresponds to a state
$|\; \underline{n} \; \rangle=
{\displaystyle\prod_{j=1}^{L}}
(c_{j}^{\dagger})^{n_{j}}|\; 0 \; \rangle$
where $|\; 0 \; \rangle$ is the vacuum containing no particles
and $L$
is the number of available lattice sites. A vector in the
Fock space is decomposed as
 $|\; F(t) \; \rangle=
{\displaystyle \sum_{\{n_{j}\}}}
F(\underline{n},t)|\; \underline{n} \; \rangle$
and a scalar product is defined by
$\langle\underline{n}|\underline{m}\rangle
={\displaystyle\prod_{j=1}^{L}[\delta_{n_{j},m_{j}}n_{j}!]}$\ .

For noninteracting classical particles
the creation operators $c_{j}^{\dagger}$
acting on site $i$
and their adjoint operators $c_{j}$
which annihilate particles obey bosonic
commutation relations \cite{doi}.
If we consider a system of particles with
exclusion, i.e., if at most one particle is allowed to be on each
lattice site,
the creation and annihilation operators fulfill Pauli type
commutation rules:
operators on the same lattice site have fermionic anticommutation
relations whereas operators for different lattice sites commute
\cite {sa1,ru}.
For simplicity we will refer to exclusive particles as fermions.

Here we are interested in correlation functions. If the physical
quantities $A(\underline{n})$ and
$B(\underline{n})$ are analytical functions of the
occupation numbers we find \cite{dis}:
\begin{equation}\label{2E}
<A(t)B(0)>=\langle\;s \; |AU_{t}B|\; F(0)\;\rangle
\end{equation}
with $\langle \; s \;|=\langle \; 0 \; |\mbox{e}^{\sum _{j} c_{j}}$
\cite{doi,sa1} and
$A,B$
are the corresponding functions of the particle number operators
$n_{j}=c^{\dagger}_{j}c_{j}$. Because of the normalization condition
\begin{equation}\label{2Y} \langle \; s \;|H=0 \end{equation}
which has to be valid for any $H$
describing a stochastic process we may insert
$U_{-t}$ between
the first two terms on the r.h.s. of Eq.
(\ref{2E}) and obtain an expression
very similar to the one in ordinary quantum mechanics. Note that averages
are linear in the probability distribution here,
while they are bilinear in the state
in quantum theory. For special problems, however,
linearity and bilinearity may coincide
(see (\ref{2K2})).

Of particular interest are two-point correlation functions of
particle numbers in a steady state. In what follows
we restrict ourselves to systems with particle conserving
dynamics, such
as diffusive systems. A two-point correlation function for an
$N$-particle-system is given by
\begin{equation}\label{2F}
G_{N}(x,y;t)=<n_{x}(t)n_{y}(0)>_{st}
\end{equation}
where a subscript $x$ denotes the label of the lattice site
corresponding
to a position $x$ in space and $n_{x}$ is the particle number in $x$.
Averaging is performed over the
stationary $N$-particle state.

The connected two-point correlation function
is defined
by \begin{equation}\label{2F1} C_{N}(x,y;t)=
<n_{x}(t)n_{y}(0)>_{st}-<n_{x}>_{st}<n_{y}>_{st}
\end{equation}
where the arguments $t$ and $0$ are dropped in the second term, because
$<n_{x}(t)>_{st}$ does not depend on time.

First we consider a system with one particle only. It can be described
using a master equation (\ref{2A}) or much easier by means of the
following probability distribution:
\begin{equation}\label{2G}
P(x,t;y,0)=P(x,t|y,0)P(y,0)
\end{equation}
which gives the probability of finding the particle in $y$ at
time $0$ and
in $x$ at time $t$. The first expression on the r.h.s.
is the corresponding
conditional probability. For the two-point correlation function defined
above we obtain \begin{eqnarray}
G_{1}(x,y;t)&=&\sum_{\{n_{j}\},\{n'_{j}\}}^{
  \;\;\;\;\; \;(1)} n_{x}n'_{y}
  F(\underline{n},t;\underline{n'},0) \nonumber\\
\label{2H} &=&P(x,t|y,0)P_{st}(y)
\end{eqnarray}
where $F(\underline{n},t;\underline{n'},0)$ is the
combined probability of
configuration $\underline{n}$ at time $t$ and
$\underline{n}$ at time $0$.
(A superscript $(k)$ at the sum means that the sum runs
over states with a total particle number $k$ only).
Assuming furthermore the steady state to be homogeneous we get:
\begin{equation}\label{2I}
G_{1}(x,y;t) =\frac{1}{L}P(x,t|y,0)
\hspace*{2mm} .
\end{equation}

Our aim is to express  $C_{N}$ in terms of $P(x,t|y,0)$.
First we will find  relations  between $G_{N}$ and  $G_{1}$ ,
one valid in
the fermionic case (with particle exclusion) and another one for bosons.
We will start with the fermionic case. The calculations for bosons are
analogous and are not shown here for sake of brevity.

\subsection{The fermionic case}
It is easy to check that in the fermionic case the operators
\begin{equation}\label{2I3}
S^{+}=\sum_{j=1}^{L}c_{j}\;\;,
\;\;S^{-}=\sum_{j=1}^{L}c^{\dagger}_{j}\;
\;,\;\;S^{z}=\sum_{j=1}^{L}(-n_{j}+\frac{1}{2})
\end{equation}
satisfy the commutation relations of $SU(2)$.

In what follows we will study systems the dynamics of which
are given by
an $SU(2)$-symmetric Hamiltonian $H$, i.e.,
\begin{eqnarray}
\label{2I1}
[H,S^{\pm}]_{-}=0\\
\label{2I2}
[H,S^{z}]_{-}=0
\hspace*{2mm} .
\end{eqnarray}
(In the discrete time case $H$ has to be replaced by the transfer
matrix $T$.)

As a consequence of this symmetry some interesting properties result:\\
(i) The number of particles in the system
is a conserved quantity because of Eq. (\ref{2I2}).\\
(ii) The Hamiltonian is hermitian.\\
Furthermore, because of the normalization condition (\ref{2Y}) and
Eq. (\ref{2I1}) we find:  \begin{eqnarray}
\label{2Z}
H | \; N \;\rangle&=&0 \;\;\; \mbox{with} \;\;\;
| \; N \;\rangle=\frac{1}{N!}(S^{-})^{N} |\; 0 \; \rangle\\
\label{2Z8}
 \mbox{and}\;\;\; \langle \; N \;|H&=&0 \;\;\; \mbox{with} \;\;\;
\langle \; N \;|=\frac{1}{N!}| \; 0 \; \rangle(S^{+})^{N}
\hspace*{2mm} .
\end{eqnarray}
Hence, the normalized N-particle state ($N<L$)
\begin{equation}\label{2Z1}
| \; N \;\rangle_{norm}=\BIN{L}{N}^{-1}
| \; N \;\rangle=\BIN{L}{N}^{-1}
\sum_{\{n_{j}\}}^{\;\;\;
\;\;\;\;(N)}| \; \underline{n} \;\rangle
\hspace*{2mm} .
\end{equation}
is a stationary solution of the problem assigning the same probability to
any possible configuration of the $n_{j}$
which has a total particle number
$N$. (The summation in (\ref{2Z1}) runs over states with $n_{j}=0,1$ and
$ {\displaystyle \sum_{j=1}^{L}n_{j}}=N$ .) The averaged occupation
numbers in state $| \; N \;\rangle_{norm}$ are $<n_{x}>=N/L$ .

{}From Eq. (\ref{2F}) and (\ref{2E}) we find for the correlation function:
\begin{equation}\label{2K1}
G_{N}^{ferm}(x,y;t)=
\langle \; s \;| n_{x}U_{t}n_{y}| \; N \;\rangle_{norm}
\end{equation}
where the steady states (\ref{2Z1}) are used.
Because of $\langle \; s \;|={\displaystyle
\sum_{N=0}^{L}}\langle \; N \;|$, particle conservation
and the orthogonality of the $| \; N \;\rangle$ we get:
\begin{equation}\label{2K2}
\BIN{L}{N}G_{N}^{ferm}(x,y;t)=
\langle \; N \;| n_{x}U_{t}n_{y}| \; N \;\rangle
\end{equation}
Note that expression (\ref{2K2}) can be understood as an
imaginary time scattering matrix of a quantum mechanical system since the
Hamiltonian is hermitian.

Using the symmetry (\ref{2I1}) the following recursion relation can be
derived from Eq. (\ref{2K2}) (see Appendix A, Eq. (\ref{AK})):
\begin{equation}\label{2M}
G_{N+1}^{ferm}(x,y;t)=\frac{L-N-1}{L-N} \frac{N+1}{N}
G_{N}^{ferm}(x,y;t)+\frac{1}{L}\frac{N+1}{L-N}
\hspace*{2mm} .
\end{equation}
{}From this an exact expression for $G_{N}$ in terms of $G_{1}$ can be
derived, which reduces in the thermodynamic limit $L,N \rightarrow \infty
,\rho =N/L=const.$ to
\begin{equation}\label{2O}
G_{N}^{ferm}(x,y;t)=N(1-\rho)G_{1}^{ferm}(x,y;t)+\rho^{2} \end{equation}
where $\rho$ is the mean density of the N-particle system.
Inserting Eq. (\ref{2I})
into Eq. (\ref{2O}) and using the definition (\ref{2F1}) of the connected
two-point correlation function we find:
\begin{equation}\label{2Q}
C_{N}^{ferm}(x,y;t)=\rho (1-\rho)P(x,t|y,0)
\hspace*{2mm} .
\end{equation}
Eq. (\ref{2Q}) is symmetric with respect to $\rho
\leftrightarrow
(1-\rho)$. This reflects the particle-hole symmetry in a system where
occupation numbers are restricted to 0 and 1 only. The amplitude of the
correlation function has its maximum at $\rho=1/2$.

Because the properties of a one-particle system are well known for many
environments, Eq. (\ref{2Q}) provides a useful tool to study
many-particle systems in any dimension provided the $SU(2)$-symmetry
holds. This will be demonstrated in the next section.

\subsection{The bosonic case}

Again we assume $H$ to satisfy Eqs. (\ref{2I1}) and (\ref{2I2}) of
the previous subsection. Note, however, that in the bosonic case
the operators $S^{\pm,z}$ (\ref{2I3})
do not form a $SU(2)$ algebra, instead
one has a harmonic oscillator algebra with
$[\, S^{+} , \, S^{-} \, ] = L$
where $L$ is the number of sites in the system.

As in the fermionic case the total particle number is conserved,
$H=H^{\dagger}$ is hermitian
and Eqs. (\ref{2Z}) and (\ref{2Z8}) result from the symmetry
(\ref{2I1}). Only the normalization constants and the
explicit form of the steady states are different:
\begin{equation}\label{2R}
| \; N \;\rangle_{norm} =
\frac{N!}{L^{N}}| \; N \;\rangle =
\frac{N!}{L^{N}}\sum_{\{n_{j}\}}^{\;\;\;\;\;(N)}\left(
\prod_{j=1}^{L}\frac{1}{n_{j}!}\right)
| \; \underline{n} \;\rangle
\hspace*{2mm} .
\end{equation}
These states correspond to homogeneous probability distributions as for
fermions. If the system is
in the state $| \; N \;\rangle_{norm}$ any configuration
with particle number N has a probability proportional to
${\displaystyle \prod_{j=1}^{L}}\frac{1}{n_{j}!}$\ .

In a way analogous to that in section 2.1 we find a recursion relation
\begin{equation}\label{2S}
G_{N+1}^{boson}(x,y;t)= \frac{N+1}{N}
G_{N}^{boson}(x,y;t)+\frac{N+1}{L^{2}}
\end{equation}
from which follows
\begin{equation}\label{2T}
G_{N}^{boson}(x,y;t)=N G_{1}^{boson}(x,y;t)+\rho^{2}
\end{equation}
in the thermodynamic limit.
Instead of Eq. (\ref{2Q}) we obtain for
the connected two-point correlation function for bosons:
\begin{equation}\label{2V}
C_{N}^{boson}(x,y,z)=\rho P(x,t|y,0)
\hspace*{2mm} .
\end{equation}
The probability on the r.h.s. is the same as in the fermionic case
because there is no distinction between fermion and boson for a single
particle system. The connected two-point correlation function
differs from that in the fermionic case
(\ref{2Q}) in the factor $(1-\rho)$ occuring in the latter one. This
gives a quantitative description of the effect caused
by the particle exclusion
for the class of systems considered here.
Of course, there is no particle-hole symmetry for the bosonic system.

\section{Many-particle diffusion in a random barrier environment}
\setcounter{equation}{0}

Applying the results of the last section we will study many-particle
diffusion in random environments which are constant
in time. The one-dimensional system of interest is assumed to be
described by a fermionic Hamiltonian (similar to \cite {pa,sa1}):
\begin{eqnarray} \label{3A}
H&=&\sum_{j}\{p_{j}[c^{\dagger}_{j+1}
c_{j}-c_{j+1}c^{\dagger}_{j+1}c^{\dagger}_{j}c_{j}]+
p_{j-1}[c^{\dagger}_{j-1}c_{j}-c_{j-1}
c^{\dagger}_{j-1}c^{\dagger}_{j}c_{j}]\}\\
\label{3A2}
&= &\frac {1}{2} \sum_{j}p_{j}
[\vec{\sigma_{j}}\vec{\sigma_{j+1}}-1]\;\;\;\mbox{with}\;\;
\vec{\sigma_{j}}=
\{c^{\dagger}_{j}+c_{j},-i(c^{\dagger}_{j}+c_{j}),1-2n_{j}\}
\hspace*{2mm} .
\end{eqnarray}
The components of $\vec{\sigma_{j}}$ are the Pauli matrices. We see
from Eq. (\ref{3A2}) that $H$ is the Hamiltonian of a generalized
Heisenberg antiferromagnet with space-dependent spin-spin coupling.
Hence the correlation functions calculated below
have a physical meaning for this quantum system (see section 2.1).

In what follows we will concentrate on the classical system described by
the evolution equation (\ref{2B}) with the Hamiltonian (\ref{3A}).
The effect of particle exclusion is reflected in the terms $c_{j\pm
1} c^{\dagger}_{j\pm 1}$ and
in the Pauli commutation relations between the
operators. The Hamiltonian of
nonexclusive (bosonic) particles moving in the
same environment is obtained from (\ref{3A}) by omitting the
aforementioned factors.

The Hamiltonian (\ref{3A}) generalizes the problems investigated in
\cite{KDN,pa} since it allows
the hopping rates $p_{j}$ to be
space dependent and stochastic. The general form (\ref{3A}) includes
classical diffusion, diffusion in structured environments and in certain
disordered media as e.g.
in the random barrier medium. Other problems such as the
random trap model are
not described by (\ref{3A}) and cannot be treated in the simple way
proposed here.

The Hamiltonian (\ref{3A}) is $SU(2)$ symmetric.
(The corresponding bosonic Hamiltonian obeys the condition from
section 2.2). Consequently, the relations (\ref{2Q}) and (\ref{2V})
between the correlator in the $N$-particle sector and the
probability distribution in the one-particle sector
are valid in
the thermodynamic limit.
For a random environment
(\ref{2Q}) and (\ref{2V}) are valid for each realization of the
environment
and $\rho=N/L$ is the same for each realization. Hence, we
can average the linear equations (\ref{2Q}) and
(\ref{2V}) over the environment (denoted by
$\overline{X}$). The result is:
\begin{eqnarray}  \label{3B}
\overline{C_{N}^{ferm}(x,y;t)}&=&\rho(1-\rho)\overline{P(x,t|y,0)}\\
\label{3B1}
\overline{C_{N}^{boson}(x,y;t)}&=&\rho\overline{P(x,t|y,0)}
\hspace*{2mm} .
\end{eqnarray}
Eqs. (\ref{3B}) and (\ref{3B1}) give the averaged two-point correlation
functions
for the many particle system provided
$\overline{P(x,t|y,0)}$ is known. The latter
quantity is the averaged solution of:
\begin{equation}\label{3C}
\partial_{t}P(x,t|y,0)=p_{x}[P(x+1,t|y,0)-P(x,t|y,0)]+
p_{x-1}[P(x-1,t|y,0)-P(x,t|y,0)]
\hspace*{2mm} .
\end{equation}
Eq. (\ref{3C}) is obviously the master equation describing the motion
of a single
particle in the environment defined by Eq. (\ref{3A}). It can be formally
derived as a correlation function (\ref{2F}) in the one particle sector
(analogous to the procedure in section 4).

We briefly review some known results.
The most simple case is that of classical diffusion, i.e., with
deterministic $p_{x}=D \;\forall x$,
for which $\overline{P(x,t|y,0)}=P(x,t|y,0)$
is well-known from textbooks \cite{ka}. For $x,t\gg 1$ it is given by:
\begin{equation}\label{3D}
P(x,t|y,0)=\frac{1}{\sqrt{4 \pi Dt}}\mbox{e}^{-\frac{x^{2}}{4Dt}}
\end{equation}
and its Fourier-Laplace transform $\tilde{S}(k,\omega)$ is
\begin{equation}\label{3E}
\tilde{S}(k,\omega) = \frac{1}{\omega + D k^2} \hspace*{2mm} .
\end{equation}
Note that $k^2$ in the denominator derives from the expansion of the
Fourier transform $2(1-\cos{k})$
of the lattice Laplacian to lowest order in $k$.
Keeping also higher terms leads to an effective frequency
dependent diffusion constant $D(k)=(1-k^2/12)D$ (see below).
Eq. (\ref{3D}) combined with (\ref{3B}) and (\ref{3B1})
gives the two-point correlation
function for the many particle system in a deterministic, homogeneous
environment of arbitrary dimension. The result for the fermionic
case in one dimension  is known from \cite{KDN,GS}.

Next we consider a random barrier environment, where the hopping rates
$p_{x}$ have the same probability distribution $g(p)$
for all sites $x$ and are not correlated.
We denote the mean value of $p$ by $\overline{p}$ and its
variance $\overline{(p-\overline{p})^2}$ by $\sigma^2$. (We shall
neglect higher moments and powers of $\sigma^2$ in all calculations
throughout this paper.) The mean value $\overline{p}=D$ is simply the
diffusion constant of an ordered system
with $g(p)=\delta(p-D)$.

For the single-particle problem
in the case of weak, uncorrelated disorder, i.e.,
if all moments $\beta_{m}=\overline{p^{-m}}$ exist and the $p_x$ are
independent random variables, the
short wavelength - low frequency behaviour of the Fourier-Laplace
transform $\overline{\tilde{S}(k,\omega)}$
 of $\overline{P(x,t|y,0)}$ is known
\cite{bo} - \cite{zw}. One introduces a generalized diffusion
constant $D(k,\omega)$ by writing
\begin{equation}\label{3F}
\overline{\tilde{S}(k,\omega)} =
\frac{1}{\omega + D(k,\omega) k^2}
\end{equation}
and finds \cite{dente,he}
\begin{eqnarray}\label{3G}
D(k,\omega) & = &
D_0 + D_1 \sqrt{\omega} + D_2 \omega + \dots \nonumber \\
 & & - k^2(E_0 + E_1 \sqrt{\omega} + \dots) + \dots
\end{eqnarray}
Here $D_j$ and $E_j$ are functions of the moments $\mu_l =
\overline{(p^{-1}-\overline{p^{-1}})^l}$
of the inverse hopping rates. Expanding them around
$D=\overline{p}$ and neglecting higher powers of $\sigma^2$
as well as higher moments $\sigma_l =\overline{(p-\overline{p})^l}$
one obtains
\begin{eqnarray}\label{3G1}
D_0 & = &  \overline{p^{-1}} \approx D - \sigma^2 D^{-1} \\
D_1 & = &  \frac{1}{2} D_0^{5/2} \mu_2
\approx \frac{1}{2}\sigma^2 D^{-3/2}
\hspace*{2mm} .
\end{eqnarray}
Within the framework of our approximation $D_2$ is 0 and
$E_0 = D_0/12, \; E_1 = D_1 /12$ and we can write
\begin{equation}\label{3H}
D(k,\omega) = D \left( 1- \frac{k^2}{12} \right)
\left( 1 - \frac{\sigma^2}{D^2}(1-\frac{1}{2} \sqrt{D\omega}) \right)
\hspace*{2mm} .
\end{equation}
The prefactor $1-k^2/12$ has its origin as above (see Eq. (\ref{3E}))
in the Fourier transform of the lattice Laplacian
and is present also in the ordered system. Up to order $\sigma^2$
the uncorrelated disorder affects only the time dependence of the
correlation function. For $x,t\rightarrow
\infty$ and $x^2/t$ fixed
the averaged probability distribution
$\overline{P(x,t|y,0)}$ has the same form as (\ref{3D}) with a
diffusion constant $D_0^{-1}=\overline{p^{-1}}$.

In the case of strong disorder (if some of the moments $\beta_{m}$ do not
exist) $\overline{P(x,t|y,0)} $ cannot be approximated by the classical
diffusion distribution \cite{bo,ha} and a phase transition
occurs. This becomes apparent in the return
probability $\overline{P(x,t|x,0)}$ \cite{ha,bo} which is
related via Eqs. (\ref{3B}) and
(\ref{3B1}) to the autocorrelation function
of the many-particle system.
Its large time behaviour is given by $\overline{P(x,t|x,0)}
\propto t^{-\alpha}$ with $\alpha=1/2$ for weak disorder and $\alpha <
1/2$ for strong disorder \cite{bo,ha}.
For more results on the random barrier
model s. \cite{bo} - \cite{he}.
A similar approach can be used for studying many-particle diffusion
in hierarchically structured media. Results for the single
particle problem are given in \cite{te,mar,dean}.

Our discussion in Sec.~2 shows that all results derived for
the probability distribution of the single particle process
are also valid for
the two-point correlation functions
for many-particle
systems given by (\ref{3B}) and (\ref{3B1}).
In the next section we are going to derive some new results
on the effect of disorder correlations.

\section{Correlation functions in a random 6-vertex model}
\setcounter{equation}{0}

In this section we restrict ourselves to one space dimension only
and consider discrete-time dynamics. It was shown by Kandel et al.
that the diagonal-to-diagonal transfer matrix of the 6-vertex
model for a certain one parameter family of vertex weights
describes diffusion
of particles with exclusion in one dimension \cite{KDN}.
In this mapping the time evolution proceeds along one diagonal
of a square lattice while the space extends along the diagonal
perpendicular to the time-diagonal (see Fig.1).
They consider the standard case where the vertex weights do not
depend on the position of the vertex in the two-dimensional
lattice. As in the preceding section we want to
study space-dependent hopping rates. In the framework of the
mapping to the vertex model on a square lattice
this is achieved by introducing
vertex weights which are constant on the time-diagonal of the lattice
but vary along the space-diagonal.
For the convenience of the reader we repeat this mapping in
Appendix B and generalize it to arbitrary, space-dependent vertex
weights.

The dynamics of the exclusion process
which leads to the generalized transfer matrix of the 6-vertex model
are defined as follows: We present the state of the system
with $L$ sites ($L$ even) at time $t$ by the quantity $\underline{n}(t)
=\{n_1(t), n_2(t),
\dots , n_L(t)\}$
where $n_x(t)$ counts the number of particles on
site $x$ and can take the values 0 or 1.
The time evolution consists of two steps. Suppose the system
is in the state $\underline{n}(t)$ with $t$ an integer.
In the first half-time step $t \rightarrow t+1/2$
we divide the chain of $L$ sites
into pairs of sites $(1,2)$, $(3,4)$, ..., $(L-1,L)$. If both
sites in a pair $(2x-1,2x)$ are occupied or empty then they remain
so with probability 1. (We exclude the possibility of particle
creation or annihilation.) If there is one particle and one hole
then the particle hops to the unoccupied site in the pair with
probability $p_{2x-1}$ and remains where it was with
probability $1-p_{2x-1}$. Note that the hopping probability in
such a pair is the same for both directions, i.e., it does not
depend on whether the particle is on site $2x-1$ or on site $2x$.
These hopping rules are applied in parallel to all pairs in the
chain. In the second half-time step $t+1/2\rightarrow t+1$
we shift the pairing of the
chain by one lattice unit such that the pairs are now $(2,3)$, ...
$(L,1)$ (we assume periodic boundary conditions). We apply the
same rules as above, but the hopping probabilities in a
pair $(2x, 2x+1)$ are now $p_{2x}$. From these rules one can
derive a master equation for the probability distribution
$F(\underline{n},t)$ (\ref{2A}).

Instead of working with a master equation, we directly study
the transfer matrix $T$
which encodes these hopping rules as discussed in section 2.
We choose as a basis of
the Fock space, the same basis as in section 3
 where the presence of a
particle corresponds to spin down and the absence of a particle
corresponds to spin up. Then the transfer matrix is given by
\begin{equation}\label{4-1}
T = T^{\rm even} T^{\rm odd} =
\prod_{j=1}^{L/2} T_{2j}(p_{2j})
\prod_{j=1}^{L/2} T_{2j-1}(p_{2j-1})
\end{equation}
where
\begin{equation}\label{4-2}
T_j(p_j) = 1 - \frac{p_j}{2} (\vec{\sigma_j}\vec{\sigma_{j+1}})
\end{equation}
with the Pauli matrices $\sigma^{x,y,z}$.
The time evolution operator $T^t$ for $t$ time steps
is defined as the $k$th power of $T$ if $t=k$ is integer
and as $T^{\rm odd} T^k$ if $t=k+1/2$.
We use periodic boundary condition and
occasionally label the spatial indices from $-L/2$ to $L/2-1$.
(Because of the periodic boundary condition they
are defined mod $L$.) The local transfer matrices
$T_j(p_j)$ act as unit operator on all sites except on the pair
$(i,i+1)$. Each $T_j$ commutes with the generators (\ref{2I3})
of $SU(2)$, so $T$ is symmetric under the action of $SU(2)$
and the time-dependent connected two-point correlation function in the
steady state
\begin{equation}\label{4-3}
C_N(x,y;t) = \langle \; s \;|
n_x T^t n_y
| \; N \;\rangle - \rho^2
\end{equation}
in the sector with $N=\rho L$ particles is given by the correlation
function $G_1(x,y;t)$ in the same environment in the one particle
sector as in
(\ref{2Q}). This is true for any choice of the $p_j$
and we can restrict our discussion to the one particle sector.
We shall omit the index 1 in the correlator and simply write
$G(x,y;t)$. In order to avoid boundary effects we shall furthermore
work in the thermodynamic limit $L \rightarrow \infty$. We denote the
lattice constants in space and time direction by $a_r$ and
$a_t$ respectively. On the square lattice one has $a_r=a_t$,
but we shall discuss later the case $a_r\neq a_t$. The integer
numbers $r=R/a_r$ and $t=\tau/(2a_t)$
measure the distance $R$ in space and $\tau$ in
time direction resp. in units of the respective lattice constants.
$a_r$ and $a_t$. Note that one full time step in the time
evolution corresponds to $\tau=2a_t$.

We denote by $| \; x \;\rangle$ the state with the particle being on site
$x$ and $\langle \; x \;|$ is
its transposed. The scalar product on this
space is given by $\langle \; x \;| \, y\, \rangle = \delta_{x,y}$
($\delta_{x,y}$ is the Kronecker symbol) and the unit operator is
$1=\sum_x | \; x \;\rangle \langle \; x \;|$.
The transfer matrix $A$ restricted to the one-particle
sector reads
\begin{equation}\label{4-4}
A = A^{\rm even} A^{\rm odd} =
\sum_{j=1}^{L/2} A_{2j}(p_{2j})
\sum_{j=1}^{L/2} A_{2j-1}(p_{2j-1})
\end{equation}
with local transfer matrices
\begin{equation}\label{4-4a}
A_{j}(p_j) = | \; j \;\rangle\langle \; j \;| +
| \; j+1 \;\rangle\langle \; j+1 \;|  - p_j
( | \; j \;\rangle - | \; j+1 \;\rangle )
( \langle \; j \;| - \langle \; j+1 \;| ) \hspace*{2mm} .
\end{equation}
The steady state with eigenvalue 1 of $A$ is the vector
$\langle\;a \; | = L^{-1/2} \sum_{x} | \; x \;\rangle$
and $| \; a \; \rangle$, which is $\langle \; s \;|$
restricted to the 1-particle sector,
is its transpose. The particle number operator
$n_x$ is simply given by
$n_x = | \; x \;\rangle \langle \; x \;|$
and the correlation function (\ref{4-3}) is the matrix element
$\langle \; x \;| A^t | \; y \;\rangle =
(A^t)_{x,y}$ of $A^t$. This power is
defined in the same way as $T^t$:
\begin{equation}\label{4-5}
A^t                            = \left\{ \begin{array}{ll}
A^k \hspace*{1cm} & \mbox{if $t = k$} \\
A^{\rm odd} A^k & \mbox{if $t = k+1/2$} \hspace*{2mm} .
\end{array} \right.
\end{equation}
With these definitions one can immediately derive a recursion
relation for $G(x,y;t)$ w.r.t. $x$ and $t$.
We have for integer values of $t=k$
\begin{equation}\label{4-5a}
G(x,y;t+1/2) =  \langle \; x \;|
A^{\rm odd} A^t  | \; y \;\rangle
\hspace*{2mm} .
\end{equation}
Inserting Eq. (\ref{4-4a}) into this expression gives
\begin{equation}\label{4-6}
G(x,y;t+1/2)  = \left\{
\begin{array}{ll}
(1-p_{x-1}) G(x,y;t) + p_{x-1} G(x-1,y;t) & \mbox{$x$ even}
\vspace*{4mm} \\
(1-p_{x}) G(x,y;t) + p_{x} G(x+1,y;t) & \mbox{$x$ odd} .
\end{array}
\right.
\end{equation}
In the same way one obtains
\begin{equation}\label{4-7}
G(x,y;t+1) = \left\{
\begin{array}{ll}
(1-p_{x}) G(x,y;t+1/2) + p_{x} G(x+1,y;t+1/2) & \mbox{$x$ even}
\vspace*{4mm} \\
(1-p_{x-1}) G(x,y;t+1/2) + p_{x-1} G(x-1,y;t+1/2) & \mbox{$x$ odd}
 \hspace*{2mm} .
\end{array} \right.
\end{equation}
{}From these recursion relations which are the analogue of the
recursion relation (\ref{3C}) together with the initial
condition
\begin{equation}\label{4-8}
G(x,y;0) = \delta_{x,y}
\end{equation}
one can compute the exact correlation function for arbitrary values
of the local hopping probabilities $p_x$.

The simplest non-trivial case is the homogeneous model $p_x = const. =
p$. In this case the system is invariant under
translations by two lattice units and
$G(x,y;t)=G(x+2z,y+2z;t)$
depends only on whether $y$ is even or odd and on
the distance $r=x-y$.
For the special choice
$p=1/2$ the correlation function was computed (in a different
way) by Kandel et al.  \cite{KDN} (see Appendix B).

We found that for arbitrary values of $p$ the
solution to the recursion relations (\ref{4-6})-
(\ref{4-7}) with initial condition (\ref{4-8}) for $x$ even and
$t$ integer is given by
\begin{eqnarray}\label{4-9}
G(x,y;t) & = &  p^{2t} \delta_{-r,2t} +
\sum_{k=1}^{  t-|r|/2}
\BIN{t-r/2}{k}
\BIN{t-1+r/2}{k-1}
p^{2t-2k}(1-p)^{2k} \hspace*{4mm} \mbox{$y$ even} \\
G(x,y;t) & = & \sum_{k=0}^{t- (|r|-1)/2}
\BIN{ t-(r-1)/2}{k}
\BIN{t+(r-1)/2}{k}
p^{2t-2k-1}(1-p)^{2k+1} \hspace*{4mm} \mbox{$y$ odd} \hspace*{2mm} .
\end{eqnarray}
For $x$ odd and $t$ integer one finds
\begin{eqnarray}\label{4-10}
G(x,y;t) & = & G(y,x;t) \hspace*{1cm} \mbox{$y$ even} \\
G(x,y;t) & = & G(y+1,x+1;t) \hspace*{1cm} \mbox{$y$ odd} \hspace*{2mm} .
\end{eqnarray}
The correlator for half-odd integer
values of $t$ is  given by relations (\ref{4-6}).
For finite densities $\rho=N/L$
one obtains the exact connected
correlation function (in the thermodynamic limit) by multiplying
the expressions (\ref{4-9}) with $\rho(1-\rho)$. Note that for
$p=1/2$ these expressions simplify considerably and we recover
the result of Ref. \cite{KDN} (see Eqs. (\ref{B-3}) -
(B.10) in Appendix B).

For large $t$ and $r$ (such that $r^2/t$ remains finite) or,
equivalently, for $a_t = a_r^2 \rightarrow 0$, one finds
from (\ref{4-9})
\begin{equation}\label{4-11}
C_N(r,t) = \rho(1-\rho) (4\pi D t)^{-1/2} \mbox{e}^{-r^2/4Dt}
\end{equation}
with the diffusion constant $D=p/(1-p)$.

Replacing $p \rightarrow p b$ and taking
the limit $b\rightarrow 0$,
one recovers the Heisenberg Hamiltonian (\ref{3A2}) as
$H=\lim_{b \rightarrow 0} b^{-1} (T-1)$. Similarly, taking the limit
$t \rightarrow \infty$ in (\ref{4-9})
such that $bt$ remains fixed leads to the correlation
function (\ref{3D})
in the continuous time formulation with $D=p$.
On the other hand,
for $p=1$ the correlation function reduces to the $\delta$-function
$\delta_{r,2t}$. Thus in this case the correlation function is
invariant under the scale transformations
$r \rightarrow \lambda r$, $t \rightarrow \lambda t$
corresponding to a dynamical exponent $z=1$. The crossover
as $p$ approaches 1 (i.e. $D \rightarrow \infty$)
will be discussed in a separate publication.

Now we study the average
behaviour of the correlation function in a random
environment. We assume all $p_x$ to be distributed in the interval
$0 \leq p_x \leq 1$ with the same distribution function $g(p_x)$
and introduce the quantity
\begin{equation}\label{4-12}
\Delta_x = p_x - \overline{p}
\end{equation}
where $\overline{p}$ denotes the mean value of the
distribution $g$.
For the sake of technical simplicity we choose as mean value
$\overline{p} =1/2$ corresponding the diffusion constant $D=1$.
As long as $\overline{p}$
is not close to 1, such a choice has no qualitative influence on
the averaged correlation function.
Furthermore we shall assume that the hopping probabilities
of even and odd lattice sites are uncorrelated,
\begin{equation}\label{4-13}
\overline{\Delta_{2x-1}\Delta_{2y}}=0
\end{equation}
while the
correlations
\begin{equation}\label{4-14}
\overline{\Delta_{2x}\Delta_{2y}}=
\overline{\Delta_{2x-1}\Delta_{2y-1}}=\sum_{\nu}h(2\nu)\delta_{r,2\nu}
\end{equation}
depend only the absolute value of the distance $r=|2y - 2x|$.
The quantity $h(0)=\sigma^2$ appearing in the sum in the
r.h.s. of Eq. (\ref{4-14}) is the variance of $\Delta_x$.
Finally, we consider only distributions which are sharply
centered around their mean value such that higher moments like
$\overline{\Delta_{2x} \Delta_{2y} \Delta_{2z}}$
etc. can be neglected
in a perturbative expansion of the averaged correlation function
in its moments.

With the definition (\ref{4-12}) of the quantities $\Delta_x$ we write
$A^{\rm odd} = A_0^{\rm odd} + \Delta^{\rm odd}$ and
analogously
$A^{\rm even} = A_0^{\rm even} + \Delta^{\rm even}$ where
$A_0^{\rm odd (even)}$ are the transfer matrices with all $p_x=1/2$.
The  averaged correlation function is averaged matrix element
\begin{equation}\label{4-15}
\overline{G(x,y;t)} = \overline{(A^t)_{x,y}} =
\langle \; x \;| \overline{\left(
A_0 + A_0^{\rm even} \Delta^{\rm odd} +
\Delta^{\rm even} A_0^{\rm odd} + \Delta^{\rm even}\Delta^{\rm odd}
\right)^t} | \; y \;\rangle \hspace*{2mm} .
\end{equation}
Neglecting all pieces with more than two Delta matrices in this
expression and using
$\overline{\Delta^{\rm even} \Delta^{\rm odd}}=0$
we obtain the lowest order correction to the correlation
function in the presence of disorder
\begin{eqnarray}\label{4-16}
\Gamma(x,y;t) & = & \overline{G(x,y;t)} - G^{(0)}(x,y;t) \nonumber \\
              & = & \sum_{l=0}^{n} \sum_{m=0}^{n-l}
\langle \; x \;| A_0^l
\overline{(\Delta^{\rm even} A_0^{\rm odd}) A_0^m
(\Delta^{\rm even} A_0^{\rm odd})} A_0^{n-l-m} | \; y \;\rangle + \\
              &   & \sum_{l=0}^{n} \sum_{m=0}^{n-l}
\langle \; x \;| A_0^l
\overline{(A_0^{\rm even} \Delta^{\rm odd}) A_0^m
(A_0^{\rm even} \Delta^{\rm odd})} A_0^{n-l-m} | \; y \;\rangle
\hspace*{2mm} . \nonumber
\end{eqnarray}
Here $G^{(0)}(x,y;t)$ denotes the correlator of the ordered system
with $p=1/2$ and $n=t-2$.
Some calculation shows that $\Gamma(x,y;t)$
can be expressed in terms of a sum of
three-point correlation functions of the ordered system:
\begin{equation}\label{4-17}
\Gamma(x,y;t) = 16 \sum_{\nu} h(2\nu) \sum_{m=0}^{n}
\left( (n-m)D_{\nu}(x,y;m) + \hat{D}_{\nu}(x,y;m)\right)
D_{\nu}(0,0;m+\mbox{\small $\frac{1}{2}$})
\end{equation}
with
\begin{eqnarray}\label{4-18}
D_{\nu}(x,y;k) & = &
\langle \; x \;| \left( A_0^{k} - A_0^{k+1} \right)
| \; y+2\nu \;\rangle \\
\hat{D}_{\nu}(x,y;k) & = &
\langle \; x \;| \left( A_0^{k+1/2} - A_0^{k+1} \right)
| \; y+2\nu \;\rangle
\hspace*{2mm} .
\end{eqnarray}
Note that through the averaging, translational invariance is
restored.
Multiplying Eq. (\ref{4-17}) by $\rho(1-\rho)$ yields
the lowest order correction to the averaged two-point
correlation function.

It is easy to compute
the autocorrelation function $\overline{G(0,0;t)}$ if
$h(2\nu) = \sigma^2 \delta_{\nu,0}$, i.e.,
in the absence of disorder correlations.
{}From Eqs. (\ref{4-17}),
(\ref{4-18}) and (\ref{B-3}) one obtains
\begin{equation}\label{4-18a}
\overline{G(0,0;t)} = \left(1+ 4\sigma^2  \frac{2t}{2t-1}
\right) G^{(0)}(0,0;t) \hspace*{2mm} .
\end{equation}
For large times this has the expected form (\ref{3G}), (\ref{3G1})
for $D=1$ and variance $4\sigma^2$:
$\overline{G(0,0;t)}\sim
(4\pi D_0t)^{-1/2}$ with $D_0= 1 -4\sigma^2$.

The behaviour of the correlation function for $r\neq 0$
and in the presence of disorder correlations becomes
more transparent after a Fourier-Laplace transformation.
The discrete
Fourier-Laplace transform (see Appendix B) of
$G^{(0)}(x,y;t)$
is given by
\begin{equation}\label{4-19}
\tilde{S}(k,\omega) =
\frac{1}{1-\mbox{e}^{- \omega}\cos^2{k}}
\end{equation}
while Fourier-Laplace transformation of
$\Gamma(x,y;t)$ yields
\begin{equation}\label{4-20}
\tilde{\Sigma}(k,\omega) =
\frac{ 4\sin^2{k}(1-\mbox{e}^{-\omega})}
{1-\mbox{e}^{-\omega}\cos^2{k}}
\left(\sigma^2
(1-\mbox{\small $\frac{1}{2}$}
\sqrt{1-\mbox{e}^{-\omega}})-
L(k,\omega) \right)
\tilde{S}(k,\omega)
\hspace*{2mm} .
\end{equation}
The function
\begin{equation}\label{4-21}
L(k,\omega) = \sqrt{1-\mbox{e}^{-\omega}}
\sum_{\nu=1}^{\infty} h(2\nu) \cos{(2ik\nu)} \left(
\frac{\mbox{e}^{-\omega/2}}{1+\sqrt{1-\mbox{e}^{-\omega}}}
 \right)^{|2\nu|}
\end{equation}
on the r.h.s. of Eq. (\ref{4-20}) is the contribution
of the disorder correlation function $h(2\nu)$ to the generalized
diffusion constant.
Eqs. (\ref{4-17}) and (\ref{4-20}) are the main results of
this section.
We would like to stress that up to this point all results are
exact first order contributions, i.e., valid for arbitrary
integer values of $r$ and $t\geq 0$ and arbitrary values of $k$ and
$\omega \geq 0$.

Now we focus on the large distance behaviour and expand
$\tilde{\Sigma}(k,\omega)$ up to order $\sqrt{\omega}$.
For uncorrelated disorder, $L(k,\omega)=0$, we recover the
result (\ref{3H}) of Refs. \cite{dente,he,zw} by expanding
$\cos{k}$ and $\exp{(-\omega)}$ to first order in their
respective arguments. In this case the generalized diffusion
constant does not depend on the frequency. This changes for
correlated disorder.
Assuming a decay of the form $h(2\nu) =
\sigma^2 \exp{(-|2\nu|/\xi)}$
one obtains for the correction
\begin{equation}\label{4-22}
L(k,\omega) \approx  \sigma^2\sqrt{\omega}
\frac{\mbox{e}^{-2(\sqrt{\omega} + \xi^{-1})}-\cos{2k}}
{\cosh{(2\sqrt{\omega} + 2 \xi^{-1})} - \cos{2k}}
\hspace*{2mm} .
\end{equation}
For $1 \ll \xi \ll \omega^{-1/2}$ this expression becomes
\begin{equation}\label{4-23}
L(k,\omega) \approx  \sigma^2\sqrt{ \omega}
\frac{\xi^{-1}}{\xi^{-2} + k^2} \rightarrow
\sigma^2 \sqrt{\omega} \xi
\hspace*{2mm} .
\end{equation}
This piece is independent of $k$ for low frequencies
$k \propto \sqrt{\omega}$
in our expansion which considers only contributions of order
$\sqrt{\omega}$.

$\omega^{-1/2}$ plays the role of  a crossover length scale
where the correlation function changes its behaviour.
For $1 \ll  \omega^{-1/2} \ll \xi$ one obtains a $k$-dependent
contribution in order $\sqrt{\omega}$:
\begin{equation}\label{4-24}
L(k,\omega) \approx  \sigma^2\sqrt{ \omega}
\frac{\sqrt{\omega}}{\omega   + k^2} = \sigma^2
\frac{\omega}{\omega   + k^2}
\hspace*{2mm} .
\end{equation}

For a decay of the disorder correlations of the form
$h(2\nu) \sim \sigma^2 |2\nu|^{-\alpha} \exp{(-|2\nu|/\xi)}$
the contribution of $L(k,\omega)$ to $D(k,\omega)$
is small for $\alpha > 1$, i.e., smaller
than of order $\sqrt{\omega}$. In this situation the contribution
of the disorder correlations can be neglected and the system
behaves as it was uncorrelated.

For $0 \leq \alpha <1$
and $\xi
\raisebox{-1.0mm}{\mbox{$\stackrel{\textstyle >}{\sim}$}}
\sqrt{\omega}$ the disorder contribution
is larger than of order $\sqrt{\omega}$ giving rise to a
qualitative change in the frequency dependence of the
diffusion constant and leading also to a $k$-dependence.
As opposed to $\alpha=0$ with infinite correlation length
the contribution still vanishes as $\omega,k^2 \rightarrow 0$.

For $h(2\nu) = \sigma^2 |2\nu|^{-1}$ (for
$\nu \geq 1$) one finds
\begin{equation}\label{4-25}
L(k,\omega) = \frac{1}{2} \sqrt{\omega}
\ln{(1+ \mbox{e}^{-2(\omega+\xi^{-1})} - 2
\mbox{e}^{-(\omega+\xi^{-1})}\cos{2k})}
\end{equation}
from which the various limiting cases can be easily derived.

Finally we
set $R=r a$ and $\tau=t a^2$ and study
the scaling limit $a \rightarrow 0$ keeping $R$ and $\tau$ fixed.
$\tilde{S}(k,\omega)$ becomes the well-known quantity
\begin{equation}\label{4-26}
\tilde{S}(k,\omega) = \frac{1}{\omega + k^2}
\end{equation}
while for the first order correction (\ref{4-20}) we obtain
\begin{equation}\label{4-27}
\tilde{\Sigma}(k,\omega) = \frac{4 k^2}
{\omega+ k^2}
\left(\sigma^2 - \sqrt{\omega}
\lim_{a\rightarrow 0} aL(ka,\omega a^2)  \right)
\tilde{S}(k,\omega) \hspace*{2mm} .
\end{equation}
{}From this expression we realize that the contribution of
disorder correlations vanishes if their correlation length is finite.
Furthermore, if they decay with a power law, $h(2\nu) \sim
|2\nu + 1|^{-\alpha}$, then its contribution still vanishes for
$\alpha \geq 0$. Only for an exponential decay $h(2\nu) \sim
|2\nu|^{-\alpha} \exp{(-|2\nu|/\xi)}$ with $\alpha=0$,
and an infinite correlation length
$\xi
\raisebox{-1.0mm}{\mbox{$\stackrel{\textstyle >}{\sim}$}}
\omega^{-1/2}$, does one obtain a finite contribution
(we study the limit $\omega \rightarrow 0$).
$\alpha =0$ means that all
fluctuating hopping rates $p_{2x}$ would be equal to some
quantity $p^{\rm even}$ and all $p_{2x-1}$ would be equal
to $p^{\rm odd}$ and the
averaged correlation function would be an average over
semihomogeneous models with different hopping rates at
even and odd time steps. On the other hand, $\alpha = \infty$
corresponds to completely uncorrelated choices of the
hopping rates.
However, periodic correlations of the type
$\sigma^2\cos{2\nu u}$ or $\sigma^2\sum_n \delta_{2\nu,n}$ or
similar aperiodic types of
behaviour also give a finite contribution to
the averaged correlation function in the infinite time limit.

Our results show that with decreasing
correlation $\tilde{\Sigma}(k,\omega)$ increases until
one reaches $\alpha=1$. For a stronger decay only the
variance $\sigma^2$ of the distribution function is relevant.

\section{Conclusion}
\setcounter{equation}{0}

We have studied systems of particles hopping stochastically on lattices
of arbitrary dimension with space-dependent hopping probabilities.
For exclusion models with SU(2)-invariant dynamics (see (\ref{2I3})-
(\ref{2I2})) the time-dependent
two-point correlation function of occupation numbers in the steady state
is shown to be proportional to
the probability distribution in space of a single particle moving
in the same environment (see Eq. (\ref{2Q})). For bosonic systems
 we obtain an analogous result
(\ref{2V}) if the time evolution operator  commutes with the
generators of the harmonic oscillator algebra.
The factor of proportionality
for fermions is different to the one for bosons.
This reflects the effect
of particle exclusion, i.e., of the hard core repulsion.
Using known results for single particle diffusion in
one-dimensional, disordered media, we obtain
expressions for many-particle correlation
functions for the same models.

The Fock space formalism we have applied reveals relations to quantum
systems. In particular, a classical diffusive system of
exclusive particles is shown
to be equivalent to a generalized Heisenberg antiferromagnet.

Focussing on one-dimensional systems we study a version of the
random barrier model with spatially correlated disorder. Its
time evolution is given by the diagonal-to-diagonal
transfer matrix of the 6-vertex model
with a certain random choice
of vertex weights. We study the steady state and
derive expressions (\ref{4-17}) and (\ref{4-20})
for the averaged
time-dependent two-point density correlation function in the
$N$-particle sector in the presence of weak disorder.
These expression are exact in the lowest order of the
expansion of the averaged function in the moments of distribution
of the hopping rates.
In the language of the two-dimensional vertex model these are
arrow-arrow correlation functions in the plane.

We compare our results with known results for systems with uncorrelated
disorder. If the correlation length $\xi$ of the disorder correlations
is finite or if the correlations $h(r)$ have infinite range but
decay faster than $r^{-1}$, the contribution of the disorder correlations
becomes negligibly small for large times and the system behaves like
the system with uncorrelated disorder. For infinite-ranged disorder
with a slower decay, $h(r) \sim
r^{-\alpha}$ where $0 \leq \alpha \leq 1$,
the correlation function changes its behaviour even after long times
and the generalized diffusion constant $D(k,\omega)$ becomes
$k$-dependent even in the lowest order of the expansion. For uncorrelated
disorder the diffusion constant depends only on $\omega$ in this
approximation. Finally we studied the infinite time limit.
For $\alpha > 0$ there is no contribution from the correlations,
but for $\alpha=0$, i.e., for a correlation that does not decay
for $r \rightarrow \infty$,
we observe a qualitatively different behaviour.

An interesting open question is the applicability of the ideas
of Section~2 to models with particles of different species.

\section*{Acknowledgments}

The authors would like to thank E. Domany, H. Patzlaff and
S. Trimper for stimulating discussions. Financial support
by the Minerva Foundation (S.S.) and the Deutsche
Forschungsgemeinschaft (G.S.) is gratefully acknowledged.

\appendix
\setcounter{section}{0}
\renewcommand{\theequation}{\Alph{section}.\arabic{equation}}
\setcounter{equation}{0}
\section{Recursion relation for the 2-point function}

Starting from Eq. (\ref{2K2}) for the N particle correlation function
and assuming the symmetry (\ref{2I1}) we will prove the recursion
relation
(\ref{2M}). \\For this we will use the following equations which can be
proofed easily:
\begin{eqnarray} \label{A1}
S^{-}| \; N \;\rangle&=&(N+1)| \; N+1 \;\rangle\\
\label{A2}
c_{j}^{\dagger}| \; N \;\rangle&=&n_{j}| \; N+1 \;\rangle\\
\label{A3}
c_{j}| \; N \;\rangle&=&(1-n_{j})| \; N-1 \;\rangle\\
\label{A4}
S^{+}| \; N \;\rangle&=&(L-N+1)| \; N-1 \;\rangle \\
\label{A5}
[n_{j},S^{-}]_{-}&=&c^{\dagger}_{j}
\hspace*{2mm} .
\end{eqnarray}
Writing down an expression similar to (\ref{2K2}) for $G_{N+1}^{ferm}$
and using Eq. (\ref{A1}) it results:
\begin{equation}\label{AI}
(N+1)\BIN{L}{N+1}G_{N+1}^{ferm}(x,y;t)=\langle \; N+1 \;|
n_{x}U_{t}n_{y}S^{-}| \; N \;\rangle
\hspace*{2mm} .
\end{equation}
By means of (i) and Eq. (\ref{A5}) we obtain:
\begin{eqnarray*}
(N+1)\BIN{L}{N+1}G_{N+1}^{ferm}(x,y;t)&=&\langle \; N+1 \;|
S^{-}n_{x}U_{t}n_{y}| \; N \;\rangle  \\
& &+\langle \; N+1 \;| c^{\dagger}_{x}U_{t}n_{y}| \; N \;\rangle\\
& &+\langle \; N+1 \;| n_{x}U_{t}c^{\dagger}_{y}| \; N \;\rangle
\hspace*{2mm} .
\end{eqnarray*}
Because of (\ref{A2}) and (\ref{A3}) and (\ref{A4}) this results in:
\begin{equation}\label{AJ}
N\BIN{L}{N+1}G_{N+1}^{ferm}(x,y;t)=
(L-N-1)\BIN{L}{N}G_{N}^{ferm}(x,y;t)+\langle \; N \;|U_{t}n_{y}
| \; N \;\rangle
\hspace*{2mm} .
\end{equation}
Because of the normalization condition (\ref{2Y})
 the second term on the r.h.s. of Eq. (\ref{AJ}) is equal to
$\frac{N}{L}\BIN{L}{N}$. Thus Eq. (\ref{AJ}) results in the
following recursion relation:
\begin{equation}\label{AK}
G_{N+1}^{ferm}(x,y;t)=\frac{L-N-1}{L-N} \frac{N+1}{N}
G_{N}(x,y;t)+\frac{1}{L}\frac{N+1}{L-N}
\hspace*{2mm} .
\end{equation}

\section{The 6-vertex model as a disordered diffusive system}
\setcounter{equation}{0}

Here we repeat the mapping of Ref. \cite{KDN} of a one-dimensional
diffusion problem to a 6-vertex model and generalize it to a version
of the random barrier model.
Consider the 6-vertex model on a diagonal square lattice defined
as follows: Place an up- or down-pointing arrow on each link of
the lattice and assign a non-zero Boltzmann weight to each of
the vertices shown in figure~1. (All other configurations of
arrows around an intersection of two lines, i.e., all other vertices,
are forbidden.) The partition function is the sum of the products
of Boltzmann weights of a lattice configuration taken over all
allowed configurations.
In the transfer matrix formalism
up- and down-pointing arrows represent the
state of the system at some given time $t$ (Fig. 1).
Each row of a
diagonal square lattice is built by $M$ of these vertices.
Corresponding to the $M$ vertices there are $L=2M$ sites in each row.
The configuration of arrows in the next row above (represented by the
upper arrows of the same vertices) then corresponds to the state of
the system at an intermediate time $t'=t+1/2$, and the
configuration after a full time step $t''=t+1$ corresponds to the
arrangement of arrows two rows above.
Therefore each vertex represents a local transition
from the state given by the lower two arrows of a vertex representing
the configuration on sites $j$ and $j+1$ at time $t$
to the state defined by the upper two arrows representing the
configuration at sites $j$ and $j+1$ at time $t+1/2$.
The correspondence of the vertex language to the particle picture
used in Section~4 can be understood by considering
 up-pointing arrows as
particles occupying the respective sites of the chain
while down-pointing arrows represent vacant sites, i.e., holes.

The diagonal-to-diagonal transfer matrix $T$
acting on a chain of $L$ sites ($L$ even)
of the six-vertex model with space dependent vertex weights
as shown in figure~1 is then defined by \cite{Vega}
\begin{equation}\label{B-1}
T = \prod_{j=1}^{L/2} T_{2j} \cdot \prod_{j=1}^{L/2} T_{2j-1} =
       T^{\rm even} \, T^{\rm odd} \hspace*{2mm} .
\end{equation}
The matrices $T_j$ act nontrivially on sites $j$ and $j+1$ in the chain,
on all other sites they act as unit operator.
All matrices $T_j$ and $T_{j'}$ with $|j-j' | \neq 1$ commute.
For an explicit representation of the transfer matrix
we choose a spin-1/2
tensor basis where the Pauli-matrix $\sigma_j^z$ acting on site
$j$ of the chain is diagonal and spin down at site $j$ represents
a particle (up-pointing arrow) and spin up a hole (down-pointing
arrow). In this basis
$n_j = \mbox{\small $\frac{1}{2}$}
(1 - \sigma_j^{z})$ is the projection operator
on particles on site $j$ and
$s_j^{\pm} = \mbox{\small $\frac{1}{2}$} ( \sigma_j^x \pm i \sigma_j^y )$
($\sigma^{x,y,z}$ being the Pauli matrices) create ($s_j^{-}$)
and annihilate ($s_j^{+}$) particles respectively.
The matrices $T_j$ in this basis are defined by
\begin{equation}\label{B-2}
T_j =
\left( \begin{array}{cccc}
       1 & 0 & 0 & 0 \\
       0 & 1-p_j & p_j & 0 \\
       0 & p_j & 1-p_j & 0 \\
       0 & 0 & 0 & 1 \end{array} \right)_{j,j+1} \hspace*{2mm} .
\end{equation}

In the particle language
the matrices $T_j$ describe the local transition probabilities
of particles moving from site $j$ to site $j+1$ represented
by the weights of the corresponding vertices.
With these identifications the vertex model with a random choice
of the numbers $p_j$ in the interval $0 \leq p_j \leq 1$
becomes a discrete-time version of the random barrier model.

The transfer matrix
acts in parallel first on all odd-even pairs of sites $(2j-1,2j)$,
then on all even-odd pairs.
In a model with transfer matrix $\tilde{T} =
T^{\rm odd} T^{\rm even}$ one would start the time evolution
at an intermediate half-odd integer time step and
there will be no  difference in the
physical properties of these two systems.
We assume periodic boundary conditions, i.e., we identify
site $L+1$ with site $1$.

The distinction between even and odd space coordinates
becomes physically meaningful in the limiting case when all
$p_j$ are close to, or equal to one. Suppose $p_j = 1 \forall j$. Then
particles which are on odd lattice sites at integer time steps
move to right at a constant rate of two lattice units in space
direction per full time step while particles move to left with
same velocity which is the velocity of light in the system.
Thus we have a system of non-interacting massless
relativistic right- and left-movers. If the $p_j$ are not equal
but close to 1, one expects the particles to remain relativistic
but massive. The homogeneous massive system is studied in detail
in \cite{grimm}. This feature of the vertex model makes it more
interesting than the continuous-time formulation by the Hamiltonian
(\ref{3A}) which allows only for nonrelativistic diffusion.
In what follows we will use the term
right(left)-movers for particles on odd (even) lattice sites
also for the nonrelativistic case.

The time-dependent connected two-point correlation function
for the homogeneous model with $p=1/2$ was computed by
Kandel et al. \cite{KDN} for full time steps in the continuum
limit $L \rightarrow \infty$.
We quote their
result and the corresponding expression for half-odd integer
time intervals which are used in Section 4. One obtains with $r=x-y$
and $\rho=N/L$:
\vspace*{4mm} \\
\noindent \underline{$t$ integer:} \vspace*{4mm} \\
\begin{eqnarray}\label{B-3}
\mbox{a) $x,y$ even} \hspace*{1cm} C_N(x,y;t) & = & \rho(1-\rho)
\left( \frac{1}{2} \right)^{2t}
\BIN{2t-1}{t+r/2} \\
\mbox{b) $x$ odd, $y$ even} \hspace*{1cm} C_N(x,y;t) & = &  \rho(1-\rho)
\left( \frac{1}{2} \right)^{2t}
\BIN{2t-1}{t+(r-1)/2} \\
\mbox{c) $x$ even, $y$ odd} \hspace*{1cm} C_N(x,y;t) & = & \rho(1-\rho)
\left( \frac{1}{2} \right)^{2t}
\BIN{2t-1}{t+(r-1)/2} \\
\mbox{d) $x,y$ odd} \hspace*{1cm} C_N(x,y;t) & = & \rho(1-\rho)
\left( \frac{1}{2} \right)^{2t}
\BIN{2t-1}{t-r/2}
\hspace*{2mm} .
\end{eqnarray}
\vspace*{4mm} \\
\noindent \underline{$t$ half-odd integer:} \vspace*{4mm} \\
\begin{eqnarray}\label{B-4}
\mbox{a) $x,y$ even} \hspace*{1cm} C_N(x,y;t) & = & \rho(1-\rho)
\left( \frac{1}{2} \right)^{2t}
\BIN{2t-1}{t+(r-1)/2} \\
\mbox{b) $x$ odd, $y$ even} \hspace*{1cm} C_N(x,y;t) & = &  \rho(1-\rho)
\left( \frac{1}{2} \right)^{2t}
\BIN{2t-1}{t+r/2} \\
\mbox{c) $x$ even, $y$ odd} \hspace*{1cm} C_N(x,y;t) & = & \rho(1-\rho)
\left( \frac{1}{2} \right)^{2t}
\BIN{2t-1}{t-r/2} \\
\mbox{d) $x,y$ odd} \hspace*{1cm} C_N(x,y;t) & = & \rho(1-\rho)
\left( \frac{1}{2} \right)^{2t}
\BIN{2t-1}{t+(r-1)/2}
\hspace*{2mm} .
\end{eqnarray}

Because of the distinction of right- and left-movers and full
and half-time steps  we have to define carefully Fourier and
Laplace transforms of space and time-dependent correlation
functions. We define the dynamic structure function
$\overline{S(k,t)}$ as the sum of Fourier transforms
of the two-point correlation function between right-movers and
left-movers:
\begin{equation}\label{B-5}
\overline{S(k,t)} = \sum_x \mbox{e}^{ikx}
(\overline{C(2x,0;t)} + \overline{C(2x+1,1;t)})
\hspace*{2mm} .
\end{equation}

{}From (\ref{B-3}) one obtains for the correlation function
$C_N(x,y;t)$ of the ordered system with $p=1/2$ at integer times
\begin{equation}\label{B-x}
S_0(k,t) =  \rho(1-\rho) \left( \cos{k} \right)^{2t}
\hspace*{2mm} .
\end{equation}
The discrete Laplace transform of a time-dependent quantity
$f(t)$ is defined as the sum
\begin{equation}\label{B-y}
\tilde{f}(\omega) = \sum_{t=0}^{\infty} \mbox{e}^{-\omega t} f(t)
\end{equation}
over full time steps only. For the dynamic structure function
(\ref{B-x})
we obtain
\begin{equation}\label{B-z}
\tilde{S}_0(k,\omega) =
\frac{1}{1-\mbox{e}^{-\omega}(\cos{k})^2} \hspace*{2mm} .
\end{equation}

\newpage
\bibliographystyle{unsrt}

\newpage

\begin{figure}
\setlength{\unitlength}{3.7mm}
\begin{center}
\begin{picture}(40, 8)
\thicklines
\put (0,3){\line(1,1){4.}}
\put (0,7){\line(1,-1){4.}}
\put (1.2,4.2){\vector(1,1){0}}
\put (3.2,6.2){\vector(1,1){0}}
\put (0.8,6.2){\vector(-1,1){0}}
\put (2.8,4.2){\vector(-1,1){0}}
\put (6,3){\line(1,1){4.}}
\put (6,7){\line(1,-1){4.}}
\put (6.8,3.8){\vector(-1,-1){0}}
\put (8.8,5.8){\vector(-1,-1){0}}
\put (9.2,3.8){\vector(1,-1){0}}
\put (7.2,5.8){\vector(1,-1){0}}
\put (14,3){\line(1,1){4.}}
\put (14,7){\line(1,-1){4.}}
\put (14.8,3.8){\vector(-1,-1){0}}
\put (16.8,5.8){\vector(-1,-1){0}}
\put (14.8,6.2){\vector(-1,1){0}}
\put (16.8,4.2){\vector(-1,1){0}}
\put (20,3){\line(1,1){4.}}
\put (20,7){\line(1,-1){4.}}
\put (21.2,4.2){\vector(1,1){0}}
\put (23.2,6.2){\vector(1,1){0}}
\put (23.2,3.8){\vector(1,-1){0}}
\put (21.2,5.8){\vector(1,-1){0}}
\put (28,3){\line(1,1){4.}}
\put (28,7){\line(1,-1){4.}}
\put (28.8,3.8){\vector(-1,-1){0}}
\put (31.2,6.2){\vector(1,1){0}}
\put (29.2,5.8){\vector(1,-1){0}}
\put (30.8,4.2){\vector(-1,1){0}}
\put (34,3){\line(1,1){4.}}
\put (34,7){\line(1,-1){4.}}
\put (36.8,5.8){\vector(-1,-1){0}}
\put (35.2,4.2){\vector(1,1){0}}
\put (37.2,3.8){\vector(1,-1){0}}
\put (34.8,6.2){\vector(-1,1){0}}
\put (1.8,0){1}
\put (7.8,0){1}
\put (15.8,0){$p$}
\put (21.8,0){$p$}
\put (29.8,0){$1-p$}
\put (35.8,0){$1-p$}
\end{picture}
\end{center}
\caption{\protect\small
Allowed vertex configurations in the six-vertex model
and their Boltzmann weights.
Up-pointing arrows correspond to particles, down-pointing
arrows represent vacant sites. In the dynamical interpretation
of the model the Boltzmann weights give
the transition probability of the state represented by the
pair of arrows below the vertex to that above the vertex.}
\end{figure}
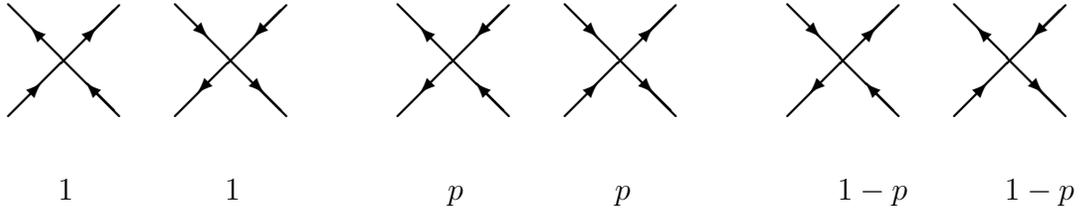

\end{document}